\documentclass[prl,floatfix,twocolumn,showpacs,amsmath,amssymb,final,superscriptaddress]{revtex4}
\usepackage{graphicx} 
\usepackage{dcolumn} 
\usepackage{bm} 
\usepackage{times}
\usepackage[pdftex,colorlinks,bookmarks=false,citecolor=blue,linkcolor=red,urlcolor=blue]{hyperref}
\usepackage{color}
\usepackage{rotating}

\begin{document}


\title{Imbalanced thee-component Fermi gas with attractive interactions: Multiple FFLO-pairing,\\ Bose-Fermi and Fermi-Fermi mixtures versus collapse and phase separation}
\author{Andreas L\"uscher}
\affiliation{Institut Romand de Recherche Num\'erique en Physique des Mat\'eriaux (IRRMA), EPFL, CH-1015 Lausanne, Switzerland}
\author{Andreas M. L\"auchli}
\affiliation{Max Planck Institut f\"ur Physik komplexer Systeme, D-01187 Dresden, Germany}

\date{\today}

\begin{abstract}
We present a detailed study of the population imbalanced three-component Hubbard chain with attractive interactions. Such a system can be realized experimentally with three different hyperfine states of ultra cold $^6$Li atoms in an optical lattice. We find that there are different phases that compete with each other in this system: A molecular superfluid phase in which the three fermion species pair up to form molecules (trions), a usual pairing phase involving two species with exactly opposite momenta, and a more exotic generalized Fulde-Ferrell-Larkin-Ovchinnikov (FFLO) phase consisting of three competing pairing tendencies with different non-zero center-of-mass momenta. At large attractive interactions the system exhibits strong tendencies towards collapse and phase separation. Employing the density-matrix-renormalization-group-method (DMRG) to determine the decay exponents of the various correlators we establish the phase diagram of this model for different fillings and interactions. 
We also discuss the experimentally relevant situation in a trap and report the existence of an additional region where two species are dynamically balanced.
\end{abstract}

\pacs{
03.75.Ss, 
03.75.Mn, 
71.10.Fd, 
71.10.Pm 
}

\maketitle

\emph{Introduction} --
Pair formation of fermions is an important phenomenon
occurring in a variety of physical systems ranging from superconductors and superfluids to dense quark matter. The pairing of {\em two} balanced 
species of fermions is well understood within the 
Bardeen-Cooper-Schrieffer (BCS) theory~\cite{bardeen57}, but pairing 
and molecule formation involving three flavors are still intriguing, especially if the Fermi surfaces of the different flavors do not match.
Recent advances in methods for trapping and controlling ultracold atoms have opened up the possibility of experimentally observing a three-component mixture of ultracold $^6$Li fermionic atoms with attractive interactions~\cite{ottenstein08,huckans08}. These experimental prospects have fueled a variety of theoretical studies of systems with equal densities of flavors, in which the formation of three-body molecules is favored for strong attraction. 
In one dimension, the molecular superfluid prevails for all attractive interactions~\cite{capponi08}, whereas in higher dimensions, an additional color superfluid phase has been found for smaller attractions~\cite{honerkamp04a}.

In the more general, but still experimentally realizable case, where the flavour densities are different, it is an open question how the system maintains its binding tendencies and whether it favors trion or pair formation. 
In the two-flavor case -- as exemplified by the imbalanced attractive Hubbard model -- an extended Fulde-Ferrell-Ovchinnikov-Larkin (FFLO) phase with a well defined momentum $Q=|k_{F\uparrow}-k_{F\downarrow}|$ of the Cooper pairs has been found~\cite{hubbardfflo}. In the three-flavor case there are three different possible FFLO 
wave vectors and it is unclear which one will dominate the pairing properties. 
Motivated by these fundamental theoretical question and their relevance 
to upcoming experiments, we have investigated the 
case of generic filling in the three-component Hubbard chain with attractive interactions - the simplest possible model describing 
these binding tendencies.
In this Letter, we present the pairing phase diagram for different
imbalance and interactions.

\begin{figure}
\centerline{\includegraphics[width=0.95\linewidth,clip]{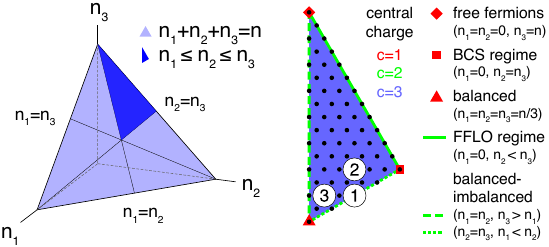}}
\caption{
\emph{(Color online) Density configurations for a given filling $n$ form a triangle in $n_1$-$n_2$-$n_3$ space. Because of symmetry, it is sufficient to consider only the subset $n_1\leq n_2 \leq n_3$, shown on the right hand side. Dots represent the systems studied in this work and the three encircled numbers refer to selected configurations presented in Figs.~\ref{fig:correlations} and \ref{fig:trap}.}
\label{fig:triangle}}
\end{figure}

\emph{Model} -- 
We concern ourselves with a population imbalanced three-component Fermi gas in a one-dimensional (1D) optical lattice, which can be described by a Hubbard Hamiltonian
\begin{equation} \label{eq:H}
H=-t \sum_{l,\alpha} \left( c_{l,\alpha}^\dag c^{\phantom{\dag}}_{l+1,\alpha} + \text{H.c.}\right)
+  \sum_{l} \sum_{\beta > \alpha}
U_{\alpha\beta} \ n_{l,\alpha} n_{l,\beta} \ .
\end{equation}
Here $c_{l,\alpha}^\dag$ ($c^{\phantom{\dag}}_{l,\alpha}$) creates (destroys) a fermion with flavor $\alpha=1,2,3$ on site $l$ and $n_{l,\alpha}= c_{l,\alpha}^\dag c^{\phantom{\dag}}_{l,\alpha}$ is the state selective occupation number operator. The parameters $t$, which we set to $t\to 1$, and
$U_{\alpha\beta} \equiv U < 0$ 
characterize the nearest-neighbor hopping and the $SU(3)$ invariant attractive on-site interaction, respectively. The flavor density configurations compatible with a given total density $n:=1/L \sum_\alpha n_\alpha$, 
where $L$ is the length of the chain, 
form a triangle in $n_1$-$n_2$-$n_3$-space. Due to permutation symmetry, it is sufficient to perform calculations on the subset $n_1 \leq n_2 \leq n_3$, see Fig.~\ref{fig:triangle}. The centroid is the point where all densities are equal. It has been shown~\cite{capponi08} that any $U<0$ leads to the formation of a Luttinger liquid of three-flavor molecules (trions). 
Along the medians, two out of the three densities are equal.
Depending on whether $n_1<n_2=n_3$ or $n_1=n_2<n_3$ holds, one encounters either a situation of flavor-1 fermions mixed with flavor (2-3) cooper pairs 
(a Bose-Fermi mixture) or trions mixed with flavor-3 fermions 
(a Fermi-Fermi mixture).
Along the edges of the triangle one recovers the physics of the two-flavor case.

{\em General considerations} --
In this 1D system
, long-ranged pairing correlations are suppressed by phase fluctuations, leading to an algebraic or exponential decay. In the case of algebraic decay, the leading asymptotic behavior of the equal-time correlators is described by power laws $C(r) \propto \cos \left(k r \right)/r^\nu$, where $k$, an appropriate integer combination of the Fermi momenta $k_{F,\alpha}$, is the ordering wave vector, and all the parameters depend on the type of fluctuation considered. The physics is determined by the dominant correlation with the slowest decay and thus the smallest exponent $\nu$. The relevant correlations for attractive interactions considered in this Letter are summarized in Tab.~\ref{tab:correlations}.
\begin{table}
	\centering
	\caption{\label{tab:correlations}
	\emph{Important correlation functions in the three-component attractive Hubbard chain.}}
	\vspace{1mm}
	\begin{ruledtabular}
		\begin{tabular}{lll}\\[-3mm]
			\multicolumn{2}{l}{Fluctuation} & Expression  \\[1mm] \hline &&\\[-2mm]
			$G_\alpha(l,l')$ & Green's functions & $\left\langle c_{l,\alpha}^{\phantom{\dag}} c_{l',\alpha}^\dag \right\rangle$ \\[2mm]
			$P_{\alpha\beta}(l,l')$ & $\alpha\beta$-pairing correlations & 
			$\left\langle c_{l,\alpha}^\dag c_{l,\beta}^\dag c_{l',\alpha}^{\phantom{\dag}} c_{l',\beta}^{\phantom{\dag}} \right\rangle $ \\[2mm]
			$T(l,l')$ & trion correlations & 
			$\left\langle c_{l,1}^\dag c_{l,2}^\dag c_{l,3}^\dag c_{l',1}^{\phantom{\dag}} c_{l',2}^{\phantom{\dag}} c_{l',3}^{\phantom{\dag}} \right\rangle$
		\end{tabular}
	\end{ruledtabular} 
\end{table}

\emph{Correlations} --
To determine the dominant fluctuations, we use the density-matrix-renormalization-group (DMRG) algorithm~\cite{White-DMRG-PRL,Schollwock-DMRG-RMP} to calculate the correlation functions indicated in Tab.~\ref{tab:correlations} for open chains with $L=128$ sites at total density $n=3/8$ and $3/4$. 
The correlation exponents are subsequently extracted by fitting the power law decays of the average correlators ${\bar C}(r)=\sum_l C(l,l+r)/(L-r)$, to reduce boundary effects.
\begin{figure}
\centerline{\includegraphics[width=0.95\linewidth,clip]{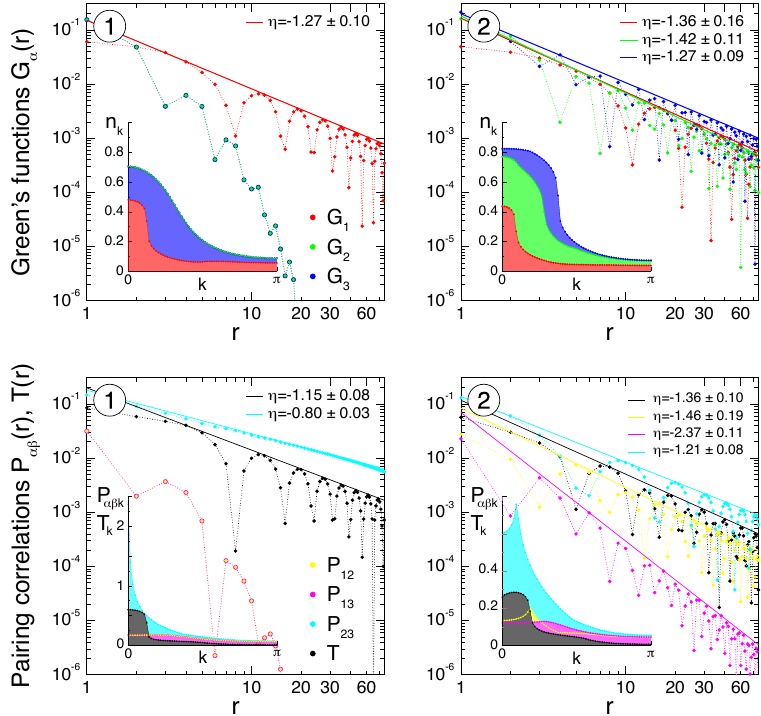}}
\caption{
\emph{(Color online) Absolute values of Green's functions and pairing correlations for two systems at density $n=3/4$ and interaction $U/t=-4$. The left column shows a balanced-imbalanced situation with ${\bf N}=(N_1,N_2,N_3)=(16,40,40)$ particles, while the right column illustrates the generic case with ${\bf N}=(12,36,48)$ fermions. The encircles numbers refer to Fig.~\ref{fig:triangle}. The inset shows the Fourier transform of the real space correlators presented in the main panels. The exponents are extracted from power law fits.} 
\label{fig:correlations}}
\end{figure}
\begin{figure*}
\centerline{\includegraphics[width=0.95\linewidth,clip]{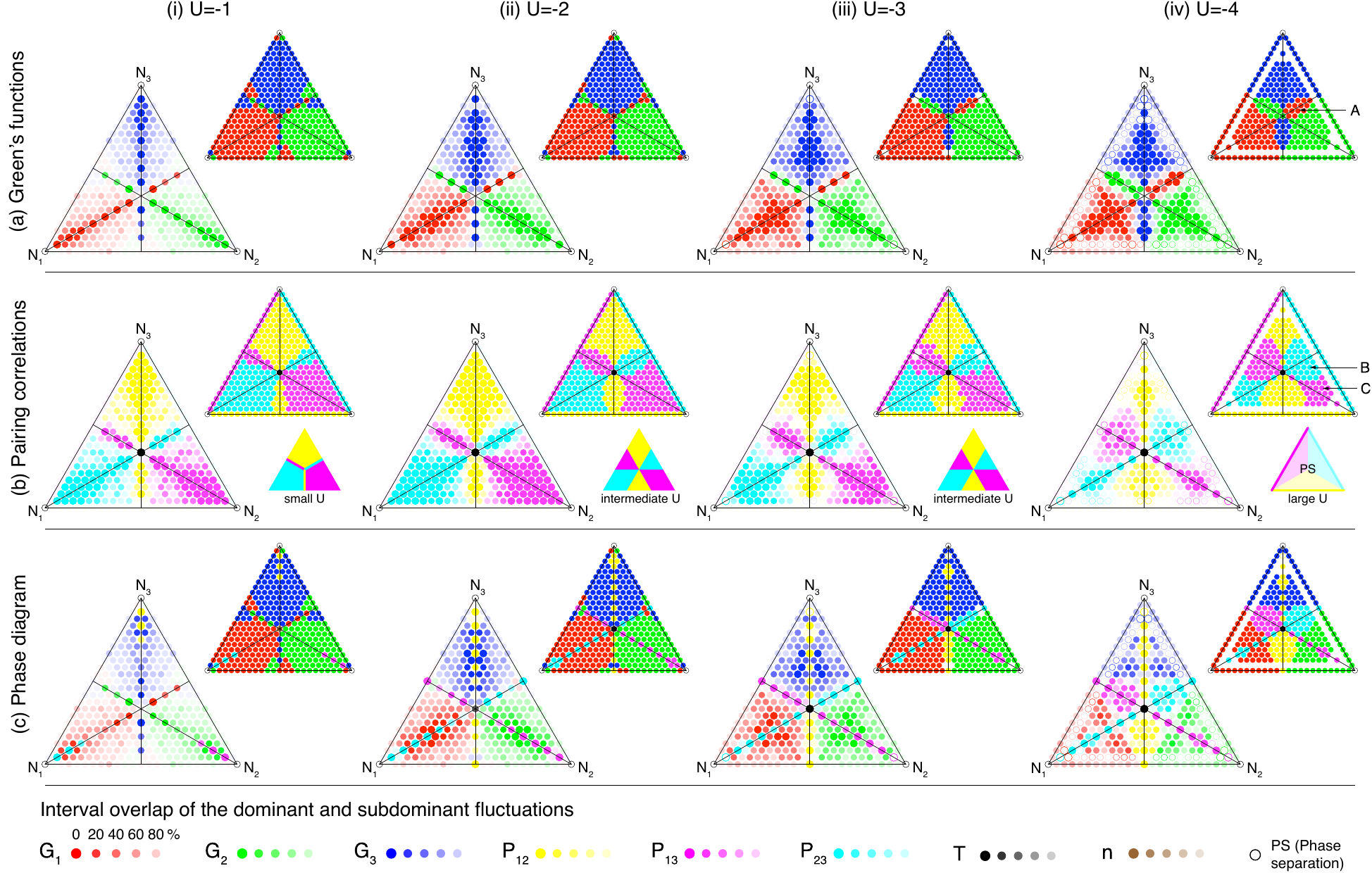}}
\caption{
\emph{(Color online) Dominant correlations of the three-component Hubbard chain at filling $n=3/4$ for various interactions $U$. The smaller triangles indicate the leading correlation function i.e. the one with the smallest decay exponent, while the larger triangles contain an additional color coding of the contrast and reliability of this determination, see text. The more intense the color, the smaller the overlap with the subdominant exponent and the more reliable the prediction. Row (a) is restricted to the decay of the three Green's functions, in row (b), the dominant pairing is indicated, while row (c) represents the combined phase diagram, including both Green's function and pairing correlation exponents.} 
\label{fig:phasediagram}}
\end{figure*}

Two representative examples for density $n=3/4$ and interaction $U/t=-4$ are shown in Fig.~\ref{fig:correlations}. In the situation depicted in the left column, two species are balanced, while the remaining flavor remains a minority ($n_1<n_2=n_3$). This corresponds to a two-component (Bose-Fermi) Luttinger liquid with expected central charge $c=2$, see Fig.~\ref{fig:triangle}. The balanced species form BCS pairs with zero momentum, as can be deduced from the strong peak at $k=0$ in the Fourier transform of the pairing correlators shown in the inset. Accordingly, single particle excitations in these two channels are gapped and the corresponding Green's functions decay exponentially. Note that the momentum of the pairing correlations of the (2-3) pairs is not affected by the presence of the spectator flavor (1). The correlation exponent is however
increased in comparison to the 
two-component model. 
Although trionic correlations also decay algebraically (here with wave vector $k_{F,1}$), they decay significantly faster than the (2-3) pairing correlations, which thus govern the physics of this regime.

If all three densities are different, the system behaves as a three-channel Luttinger liquid with expected central charge $c=3$ in which all Green's functions and pairing correlators decay as power laws. An example of such a system is shown in the right column of Fig.~\ref{fig:correlations}. Similarly to the imbalanced two-component Hubbard model~\cite{hubbardfflo}, the system now forms Fulde-Ferrell-Larkin-Ovchinnikov (FFLO) pairs with {\em finite} center-of-mass momentum. The important difference however is that 
here, there are three possible
FFLO pairs with momenta $Q^{\alpha\beta}=\left|k_{F,\alpha}-k_{F,\beta}\right|$ and correlation exponents 
$\nu^{\alpha\beta}$. This is nicely illustrated by the distinct peaks at $k=Q^{\alpha\beta} \ne 0$ in the Fourier transform of the three pairing correlations.
In this example, the pairing in channel (2-3) is dominant, pairs involving species (1-2) are subdominant, and the (1-3) pairing channel decays very quickly.
This sequence suggests an increase of the correlation exponent with increasing pairing momentum. 
Similar to the Bose-Fermi case discussed before, the molecular superfluid correlations decay more quickly than the dominant FFLO fluctuations, but they appear to decay
more slowly than the second pairing channel. Multiple FFLO wave vectors have recently been reported also for two-chain systems~\cite{feiguin:076403}.

\emph{Phase diagram} --
The complete numerical pairing phase diagram for total density $n=3/4$ and weak to moderate interactions $U/t\leq 4$ is presented in Fig.~\ref{fig:phasediagram}. Each row (a)-(c)
contains a small triangle highlighting the dominant correlation function, while the larger triangles quantify the dominance based on the fitted exponents and their uncertainty~\footnote{{To determine a confidence interval of the extracted exponents, we apply a bootstrap-like fitting procedure in which we allow the correlations functions to fluctuate within a standard deviation of the mean ${\bar C}(r)$ and restrict the fitting interval to a region $[r_\text{min},L/2]$, with variable $r_\text{min}$. This is appropriate given the fact that the power law only describes the asymptotic decay. In that way, we obtain an interval $[\Delta_\text{min},\Delta_\text{max}]$ for every exponent $\Delta$, which in turn allows us to indicate the dominant fluctuations together with a color coding of the reliability of their determination: The more intense the color, the smaller the overlap with the subdominant exponent and the more reliable the prediction.}}.
Let us first focus on the Green's functions shown in row (a) of Fig.~\ref{fig:phasediagram}. For small interactions, our results clearly reveal that the Green's functions of the majority species decay most slowly. 
With increasing $U$, one observes the growth of an additional region centered around the lower part of the median (region A in Fig.~\ref{fig:phasediagram}, indicated in the upper rightmost triangle). In this area, most of the majority species are paired up due to strong attraction. Since the density of the remaining free fermions is smaller than the density of the nominal minority flavor, the Green's function of the latter one has the slowest power law decay. 
In row (b) of Fig.~\ref{fig:phasediagram}, we present the dominant pairing correlations. The formation of trions is only favored at the SU(3) symmetric centroid.
For generic density configurations and weak interactions, we find dominant FFLO pairs formed by the minority species, except along the medians, where BCS ($Q=0$) pairing is clearly dominant. In the weak coupling limit $\left | U \right | \to 0$, we thus expect the qualitative phase diagram sketched in Fig.~\ref{fig:phasediagram}[row (b), column (i)]. In this limit, it is energetically favorable to have as many free fermions as possible, hence only the minority species form pairs. For stronger interactions, the patches around the bases of the medians (region B) spread out, hereby reducing the areas at the top (region C). 
Based on this observation, one would conjecture that in the strong coupling limit $\left | U \right | \to \infty$, majority pairs dominate, as sketched schematically in Fig.~\ref{fig:phasediagram}[(b), (iv)]. 
However, based on a perturbative argument in $t/U$, one realizes that the hopping of trions, pairs, and individual fermions intervene at third, second,
and first order, respectively. This tunable hopping asymmetry leads to different sorts of collapses, demixing, and phase separation at large $|U|/t$, reminiscent of the hopping asymmetric two flavor Hubbard 
model~\cite{batrouni-2008}. In addition, we find 
evidence for the formation of multi-particle composite liquids close to phase separation, as advocated recently in~\cite{Burovsky-09}. 
At intermediate couplings, the boundaries of the dominant correlations are surprisingly well described by lines of equal density difference, as sketched in Figs.~\ref{fig:phasediagram}[(b), (ii) and (iii)]. 
Trionic fluctuations are never dominant away from the centroid, but their correlation exponent decreases substantially with increasing attraction.

\emph{Dimensional crossover} --
Following the standard analysis of the dimensional crossover, see, e.g., Ref.~\onlinecite{giamarchi03}, we compare the dominant two-particle correlations with the single particle Green's functions and argue that upon lowering the temperature of the system, one either enters a phase dominated by the physics of the single chain FFLO correlations if the two-particle correlations decay more slowly than the Green's function, or, one first flows to a higher dimensional Fermi liquid if the 
opposite condition is satisfied.
Combining rows (a) and (b) leads to the complete phase diagram shown in row (c). For weak attraction, the fermion Green's functions are dominant, except for the centroid and the
midpoint of the edges.
Pairing correlations become more important with increasing $U$, first only along the medians, where 
$k=0$ Cooper pairs are formed, but eventually, one also observes FFLO pairing around the center of the triangle. From this sequence of images, we conclude that exotic pairing tendencies carry over to the quasi-1D case in situations where all three species have similar densities. We have  
also undertaken the same analysis for total density $n=3/8$ (not shown) and found a very similar behavior, albeit with a crossover from the Fermi liquid to the pairing phase taking place at weaker interactions. 

\emph{Influence of the trap} --
In experiments, the atomic cloud is confined in a harmonic trap of the form
\begin{equation}
H_\text{trap} = V \sum_{l=1}^L \left(l-\frac{L+1}{2}\right)^2 n_i \ .
\end{equation}
with a trapping potential V.
\begin{figure}
\centerline{\includegraphics[width=0.95\linewidth,clip]{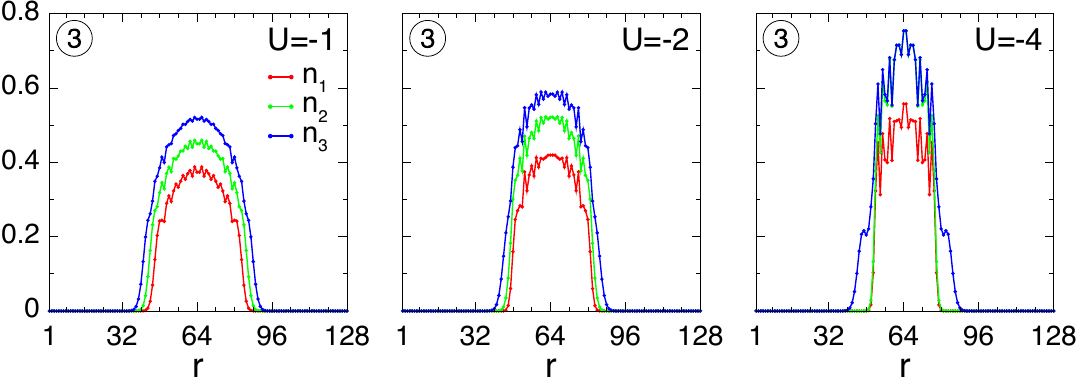}}
\caption{
\emph{(Color online) Local densities for the three-component Hubbard chain at filling $n=3/8$, with $(N_1,N_2,N_3)=(12,16,20)$ in a trap $V_\text{trap}=0.002$ for various interactions $U$. The encircled number refers to the density configuration indicated in Fig.~\ref{fig:triangle}.
} 
\label{fig:trap}}
\end{figure}
For the case of three equal flavor densities an atomic density wave has been reported in~\cite{Molina08}. For the flavor imbalanced case
an interesting effect of the trap is illustrated in Fig.~\ref{fig:trap}, for a particular filling and realistic values of $U$ and $V$. For small $U$, the area around the center of the trap resembles the corresponding homogeneous system. 
 We thus expect to find the multiple FFLO behavior discussed for the homogeneous system also in the presence of a trap. However, for stronger interactions, the system rearranges the particles in such a way that two flavors have equal densities at the center of the trap, hereby forming polarized wings. The pairing behavior remains, but the FFLO signatures disappear and one finds BCS pairs with zero momentum instead. The phase diagram presented in Fig.~\ref{fig:phasediagram} thus becomes even richer after the inclusion of the trap, exhibiting an additional BCS dominated strong coupling regime. 
At higher filling, the trap can also lead to the formation of inert inner shells, in which the density of the majority flavor is exactly equal to one. A similar effect has been observed in the two-component Hubbard chain~\cite{hubbardfflo}. 

\emph{Conclusion} -- 
We have established the pairing phase diagram of the three-component Hubbard chain with attractive interactions using the density-matrix-renormalization-group algorithm. At generic imbalance there is a coexistence of multiple FFLO modes. It will be interesting to investigate the fate of this coexistence
in higher dimensions, where Ward identities predict phase separated mixtures of paired states~\cite{cherng07}.
The phase diagram of the present non-integrable lattice model is richer than the one reported recently for an integrable three flavor continuum model~\cite{guan08}, due to the formation of multi-particle composites, collapse and phase separation at strong attraction. We are presently investigating the precise phase boundaries of these density instabilities.

\acknowledgments
We thank E.~Burovsky, S.~Capponi, M.~Haque, F.~Heidrich-Meisner, C.~Honerkamp, G.~Klingschat and G.~Roux for discussions.
This work was supported by the Swiss National Science Foundation. 

\end{document}